# Quantum-Hall physics and three dimensions

Johannes Gooth[1,2#], Stanislaw Galeski[1,2◊], Tobias Meng[3]*

[1] *Physikalisches Institut, Rheinische Friedrich-Wilhelms-Universität, Nußalle 12, 53115 Bonn, Germany*

[2] *Max Planck Institute for Chemical Physics of Solids, Nöthnitzer Straße 40, 01187 Dresden, Germany.*

[3] *Institute of Theoretical Physics and Würzburg-Dresden Cluster of Excellence ct.qmat, Technische Universität Dresden, 01069 Dresden, Germany*

Corresponding authors:

*tobias.meng@tu-dresden.de, #jgooth@uni-bonn.de, ◊stanislaw.galeski@cpfs.mpg.de

**Abstract**

The discovery of the quantum Hall effect (QHE) in 1980 marked a turning point in condensed matter physics: given appropriate experimental conditions, the Hall conductivity $\sigma_{xy}$ of a two-dimensional (2D) electron system is exactly quantized. But what happens to the QHE in three dimensions (3D)? Experiments over the past 40 years showed that some of the remarkable physics of the QHE, in particular plateau-like Hall conductivities $\sigma_{xy}$ accompanied by minima in the longitudinal resistivity $\rho_{xx}$, can also be found in 3D materials. However, since typically $\rho_{xx}$ remains finite and a quantitative relation between $\sigma_{xy}$ and the conductance quantum $e^2/h$ could not be established, the role of quantum Hall physics in 3D remains unsettled. Following a recent series of exciting experiments, the QHE in 3D has now returned to the centre stage. Here, we summarize the leap in understanding of 3D matter in magnetic fields emerging from these experiments.

**Introduction**

When a 2D electron system is exposed to strong magnetic fields $\vec{B}$ (Fig. 1), its longitudinal resistance $R_{xx}$ vanishes and its Hall conductance $G_{xy}$ is precisely given by a multiple of the fundamental constants $e^2/h$ (where $e$ is the elementary charge and $h$ is Planck's constant). This is the quantum Hall effect (QHE).[1,2] The analysis of this precise quantization of $G_{xy}$, present even in macroscopic and disordered samples, has elucidated many fundamental aspects of quantum physics, and has deepened our understanding of interacting electron systems. A new international system of units based on fundamental constants has been established, and topology has taken a central role in condensed matter physics. However, despite enormous theoretical and experimental effort, one central question has remained unsettled:

*How does the QHE generalize to three dimensions?*

Already in 1983, Avron et al.[3] proved that *gapped* electron systems in higher dimensions exhibit a quantized Hall response. More precisely, a $d$-dimensional system is characterized by $d(d-1)/2$ independent Hall conductances, each of which corresponds to a topological invariant. For a 3D electron system, those are $\sigma_{xy}$, $\sigma_{yz}$, and $\sigma_{zx}$. Specializing to a simple cubic system with lattice constant $a$, the Hall response can be cast into the form $(\sigma_{yz}, \sigma_{zx}, \sigma_{xy}) = -\frac{e^2}{ha}(m_x, m_y, m_z)$, where the integers $m_i$ satisfy a Diophantine equation.[4] Alternatively, the Hall conductivity tensor can be expressed as $\sigma_{ij} = \frac{e^2}{h}\epsilon^{ijk}\frac{G_k}{2\pi}$, where $\vec{G} = n_a\vec{G}_a + n_b\vec{G}_b + n_c\vec{G}_c$ is a reciprocal lattice vector ($\vec{G}_i$ are the primitive reciprocal lattice vectors, $n_i$ are integers).[5–7] The $n_i$ have been connected to Berry phases and Chern numbers[8], and thus equivalently generalize the topological invariant identified by Thouless, Kohmoto, Nightingale, an de den Nijs (TKNN) for the 2D case.[9]

The observation by Avron et al.[3] naturally raises the question of how gaps develop in three-dimensional electron systems subject to magnetic fields. The simplest scenario for a full bulk gap is a stack of quantum Hall layers parallel to the $(x, y)$-plane filled by spin-polarized electrons (Fig. 2). If the inter-layer tunneling is much smaller than the gaps in the individual layers, the Landau levels form weakly dispersive Landau bands (LBs), but the stack will remain gapped.[10] The stack's total Hall response is $\sigma_{xy} = \frac{e^2}{h}\frac{1}{a_z}$, where $a_z$ is the lattice constant along the $z$-direction. Due to the subleading tunneling along $z$, however, this first scenario is not truly 3D but rather quasi-2D. If the tunneling between the layers is increased, the system becomes a metal,[11] and the quantized Hall effect is destroyed.

Starting from 3D metals, Halperin[5] proposed an alternative pathway towards a full bulk: the formation of a density wave, for example be a charge density wave (CDW) or a spin density wave (SDW). Consider a 3D electron system in the quantum limit (i.e. the applied magnetic field is so strong that only the lowest Landau band crosses the chemical potential). If now a density wave forms such that its wave vector $\vec{Q} = Q\,\vec{e}_z$ spans the Fermi surface, Halperin showed that a full bulk gap develops, and that the Hall conductance is $\sigma_{xy} = \frac{e^2}{h}\frac{Q}{2\pi}$.

A density wave is, however, not the only way to gap a 3D electron system subject to magnetic fields. One has to look no further than the generalization of Hofstadter's butterfly to electrons hopping on a 3D lattice threaded by magnetic fluxes.[12,13] Provided the lattice is sufficiently anisotropic,[14,15] or if the magnetic field is not applied along a high-symmetry direction,[4] such 3D lattices exhibit a variant of Hofstadter's butterfly. Their 3D butterfly is connected to its 2D cousin, but specific to 3D.[14] As described above, the Hall responses $\sigma_{xy}$, $\sigma_{yz}$, and $\sigma_{zx}$ take quantized values that can be determined from a Diophantine equation in each of the gapped regions of the 3D butterfly.

These scenarios for a quantized 3D Hall response also predict a vanishing longitudinal resistance. Only the combined observation of quantized Hall responses and vanishing longitudinal resistances thus provides definite evidence of a truly quantized 3D Hall state. Detecting this combination of hallmarks in experiments has unfortunately proven a tough challenge for decades. Recently, the quest for 3D quantum Hall physics has been refueled by a number of studies in the strongly anisotropic 3D compounds ZrTe$_5$[16–18] and HfTe$_5$,[19,20] reporting both evidence for truly quantized Hall states resulting from a charge density wave on the one hand,[16,20] and of states with finite longitudinal resistances and no density wave on the other hand.[17,18] This ongoing debate brings one question to the center stage: can a gapless 3D electron system exhibit quasi-quantized Hall responses? If so, how could it be understood from the viewpoint of topology, and what would be its edge physics?

These questions directly relate to the long and puzzling search for quantum Hall physics in 3D and quasi-2D materials, including semiconductor multilayer superlattices,[21] organic Bechgaard salts,[22] $\eta$-$Mo_4O_{11}$,[23] $n$-doped $Bi_2Se_3$,[24] graphite,[25–28] and $EuMnBi_2$[29]. While all of these systems show plateaus in $\sigma_{xy}$ accompanied by minima in $\rho_{xx}$, the essential signatures for the multilayer QHE - a *quantized* $\sigma_{xy}$ and a *vanishing* longitudinal resistivity $\rho_{xx}$—have not been observed in any of them. Similarly, plateau-like features in $\sigma_{xy}$ accompanied by minima in $\rho_{xx}$ have been observed in truly 3D materials, such as InAs[30] and InSb[31–33], or semimetals such as Bi,[34] NbP,[35,36] HgSe[37] as well as high-$T_C$ superconductors such as YBCO.[38] While $\rho_{xx}$ in these materials can be explained in terms of dispersive 3D LBs and the Shubnikov-de Haas (SdH) effect, the observed $\sigma_{xy}$ are reminiscent of the QHE are quantitatively not understood to date. Here, we provide a new perspective on this long-standing puzzle that elegantly explains the observed conductivities on the basis of the recent experiments.

*Our main statement is that quasi-quantized plateaus along with finite longitudinal resistances are natural features of 3D metals with low charge-carrier densities subject to strong magnetic fields. While not enjoying the same topological protection as their fully-quantized cousins, these states seem rather common in materials, and – as a general rule – exhibit Hall conductivities approximately given by $\frac{e^2}{\pi h}$ times the Fermi wave vector along the direction of the magnetic field.*

We derive this hypothesis in simple terms, and connect it to the topological perspective on Hall states. This explains in which sense quasi-quantized states are connected with the historically much more studied, but experimentally apparently much less common fully-quantized states. We highlight recent experiments on semimetals and weakly doped semiconductors that alongside the historic body of measurements corroborate the proposed picture, and close this perspective with a discussion of the broader impact of our considerations.

## Theoretical considerations

Quite remarkably, the key aspects of the experimentally observed physics, namely a plateau-like Hall response $\sigma_{xy}$ at finite $\rho_{xx}$, can be understood in a rather simple model (Box 1 and Fig. 3). This model can be approached from two limits. On the one hand, one can start from a stack of decoupled quantum Hall layers, which would have perfectly flat Landau bands. Increasing the inter-layer tunneling, the LBs disperse parallel to the stacking direction (i.e. parallel to the field), and the system eventually becomes metallic when the LBs overlap.[11] On the other hand, a magnetic field can be applied to a 3D metallic system with a closed ellipsoidal Fermi surface. The electronic motion perpendicular to $\vec{B}$ is then quenched, but the electronic states still disperse parallel to $\vec{B}$, thus forming metallic Landau bands. Both viewpoints equivalently lead to the well-known conclusion that 3D electron systems subject to magnetic fields typically feature LBs that intersect the Fermi level, i.e. form a gapless state. This gapless state naturally exhibits a finite $\rho_{xx}$ whose value depends on non-universal details such as a scattering time, in agreement with the experimental findings.

Much less appreciated is the fact that that such systems can still exhibit a quasi-quantized Hall response with an amplitude set by Fermi surface properties. Let us for concreteness orient the magnetic field along the z-direction. The Hall response $\sigma_{xy}$ can then be obtained by summing the Hall conductivities for all momenta parallel to the field, $\sigma_{xy} = \frac{1}{2\pi}\int_{-\pi/a_z}^{+\pi/a_z} dk_z \; \sigma_{xy}(k_z)$, where $a_z$ is the lattice spacing along z. For fixed $k_z$, the problem effectively reduces to 2D. The Hall response can thus famously be connected to an integral over the Berry curvature. With a bit of algebra, one finds that the total Hall conductivity of a 3D metal takes the form of a sum over all occupied LBs and wavenumbers $k_z$ parallel to the field, $\sigma_{xy} = e^2/h \cdot \sum_{j\;occ} 2k_{F,z,j}$, where $k_{F,z,j}$ is the Fermi wavenumber in the magnetic field direction of the $j$th LB. The $k_{F,j}$ will in general be different for different LBs $j$. In the quantum limit, however, the 3D Hall conductivity simplifies to $\sigma_{xy} = e^2/h \cdot k_{Fz,0}/\pi$. This value can be interpreted as resulting from the limit of zero gap for a density wave of wave vector $\vec{Q} = 2\, k_{Fz,0}\, \vec{e}_z$ spanning the Fermi surface, and in this sense connects to the earlier works relating 3D Hall responses and Diophantine equations. [4,6,13,14]

Whether or not a 3D metal exhibits a plateau-like Hall response in its quantum limit sensitively depends on how $k_{Fz,0}$ shift with magnetic field. One option, usually realized in metals with large Fermi surfaces, is that the chemical potential and $k_{Fz,0}$ change in order to keep the electronic density constant. The Hall response in a single-pocket system then shows a featureless $1/B$-behavior, in stark contrast to the experiments discussed above.

Turning this argument around, any plateau-like Hall response in a single-pocket system, even in the presence of a charge or spin density wave gap, implies a field-dependent electron density.[39] Importantly, the relevant electron density here is the one of the conduction band producing the measured Hall response, and not necessarily the global electron density. This means that electron transfer between, e.g., the conduction band and localized states can change the picture. Similarly, a weakly scattering conduction band may dominate the measurable transport, while a strongly scattering additional band, possibly in parts thermally activated[40], may acts as an effective electron reservoir. Also the metallic contact leads necessarily attached to the Hall samples can serve as such a reservoir.[41] In all of these scenarios, the electron density in the conduction band can generically vary with field. This destroys the smooth $1/B$-scaling of $\sigma_{xy}$, and instead leads to the formation of plateau-like structures. Exactly how flat these plateau-like features are depending on material and sample specifics. This is where the recent experiments on topological semimetals become important. $ZrTe_5$ and $HfTe_5$ can for example be understood as weakly gapped Dirac semimetals, with the gap driving the material either into a weak or strong topological insulator phase.[42] Since the chemical potential is above the gap, however, experimental samples are typically low-density metals. As is well-known, the zeroth Landau levels in 3D Weyl and Dirac semimetals do not shift with magnetic field. An approximately constant value of $k_{Fz,0}$ (corresponding to a weakly field-dependent chemical potential in the conduction band) can therefore be reached much more easily in these systems than in metallic materials with a parabolic dispersion.

Overall, the simple picture of gapless 3D Landau bands and a weakly field-dependent Fermi level readily explains the experimentally observed quasi-constant Hall plateau in the quantum limit along with a *finite* $\rho_{xx}(B)$ for all magnetic fields. Its experimental distinction from a fully gapped quantum Hall state mainly hinges on the finite longitudinal resistivity. It is important to keep in mind that many other features are shared between quasi-quantized and fully quantized states (variable electron density,[39] importance of localized states[43]). This can experimentally blur their distinction, and Hall physics in 3D should therefore be analyzed with extreme care. Before turning to experimental considerations, however, let us first further characterize quasi-quantized states from the theory viewpoint.

***Topological perspective on 3D Hall states, and their surfaces states.***

Unlike fully gapped 3D quantum Hall states, the Hall response of 3D metals in strong $\vec{B}$ is not topologically protected. This is apparent from the fact that quasi-quantized Hall plateaus are neither strictly flat, nor exactly quantized in height.

Nevertheless, the fact that the Hall response is connected to an integral over Berry curvatures implies important similarities between metallic 3D Hall systems and the 2D QHE. In particular, we expect 2D surface states to appear parallel to $\vec{B}$, in analogy to the 1D edge channels in the 2D QHE. To show this, recall that the LB-states do not disperse perpendicular to $\vec{B} = B\,\vec{e}_z$. For given $k_z$, these states are hence either fully occupied, in which case they can be understood in terms of a filled Landau level, or else they are fully empty. For $k_z$ in the occupied momentum range, the 2D bulk-boundary correspondence implies the presence of surface states. With this in mind, consider for example the quantum limit with $\sigma_{xy} = \frac{e^2}{h}\frac{k_{Fz,0}}{\pi}$. Given that the momentum spacing along z is $\Delta k_z = \frac{2\pi}{L_z}$, there are $\frac{k_{Fz,0}}{\pi} L_z$ gapless states forming at the edge, where $L_z$ is the system length along z. This is similar to the physics described in Ref. [10] and [44], but in a gapless setting. Still, as has been discussed recently in the context of Fermi arc surface states in Weyl semimetals, Chern numbers can also be defined in globally gapless three-dimensional systems when considering gapped submanifolds in the Brillouin zone.[45] While this in principle connects the expected surface states to topology, the physics is more complex in the case of gapless 3D Hall states because the topological surface states will in general be near (in momentum space) to gapless bulk states. A more realistic model with disorder coupling bulk and edge states will therefore delocalize the edge states at least to a certain degree. Along similar lines, bulk-edge scattering will generically yield important corrections to residual edge transport in quasi-quantized Hall systems.

Indications of stacked 2D Hall effect surface states in 3D samples have already been observed in transport experiments on semiconductor multilayers parallel to the stacking direction and the magnetic field.[46] Similar experiments, combined with additional non-local transport measurements, would allow to gather evidence for the surface states of 3D Hall states in strong $\vec{B}$. These include for example local optical conductivity measurements[47] or multi-terminal voltage-probe experiments.[48,49] However, the scattering between gapless bulk and edge states implies a need for thorough theoretical modeling in order to clearly identify fingerprints of surface transport in non-local measurements.

*Scaling relations and the 3D Hall effects.*

For the 2D quantum Hall effect, the striking relation $\rho_{xx} \sim B\, d\rho_{xy}/dB$ has been identified between the diagonal and off-diagonal components of resistivity tensor.[50,51] A similar relation also holds for the components of the thermopower tensor $S_{xy} \sim B\, dS_{xx}/dB$.[52] These derivative relations can be proven for a 2D electron gas assuming local variations of $\rho_{xy}$ and $S_{xx}$ due to presence of disorder that is correlated on multiple length scales.[53] Our recent experiments have revealed that similar scaling relations hold for the $xx$ and $xy$ components of the resistivity and thermopower tensors in the 3D metals ZrTe$_5$[33], HfTe$_5$[34] and InAs[54]. However, disorder should have a less profound effect in 3D than in 2D, and the derivative relations have not yet been thoroughly understood in 3D systems. Understanding the necessary conditions for their observation in 3D thus constitutes an important open question to theory.

Moreover, Landau level crossings with $E_F$ in the 2D QHE are characterized by critical scaling of $\rho_{xx}$ and the derivative of $\rho_{xy}$ with the parameter $|B - B_c|T^{-1/\zeta}$.[55] The critical scaling is deeply rooted in disorder physics.[56,57] It is an intriguing perspective that a similar scaling might also be observable in 3D Hall systems. Future theoretical work is also necessary here to thoroughly model such disorder effects in 3D, in particular because the gapless density of states of the Landau bands has an important $B$-dependence that also affects $\rho_{xx}$.

*Correlated states in 3D materials in the quantum limit.*

While the FQHE is firmly rooted in 2D, interacting 3D electron systems in strong $\vec{B}$ can in principle also exhibit fractional states.[58–60] The simplest example of such a fractional state in 3D is a stack of decoupled Laughlin states.[10] The next level of complexity are states with weak interlayer coupling, but fixed particle number in each layer.[60–64] An even more general construction for fractional states in 3D are multilayer fractional quantum Hall states supporting gapped quasiparticles with fractional charge propagating freely in 3D.[65] These states are in principle allowed for strong Coulomb energies. Intriguingly, hints towards correlated phases of yet unknown character have recently been reported in the quantum limit of graphite,[25] ZrTe$_5$,[16] and HfTe$_5$.[19] However, the relation of these experiments to the predicted generalization of the FQHE in three dimensions is unclear, in parts because the signatures where only investigated in $\rho_{xx}$, or because the filling factor could not be quantitatively determined.

# Experimental considerations

*Low-density materials as the most promising material candidates.*

Quite generally, the cleanest experimental signatures of quantum Hall physics arise when only the lowest few Landau levels (in 2D) or Landau bands (in 3D) cross the Fermi level. In normal metals with large Fermi surfaces, such as copper, this requires inaccessibly large magnetic fields $B > 10{,}000$ T. One can therefore not expect to observe fully or quasi-quantized Hall responses in these materials. Instead, the experimental focus shifted towards semimetals and slightly doped semiconductors exhibiting small Fermi pockets. In these materials, the quantum limit can be reached in experimentally accessible magnetic fields $B < 70$ T. Given the recent surge in topological material science, in which the growth of semimetallic single crystals plays a key role,[66] it is natural that also the question of 3D Hall physics has resurfaced now.

Furthermore, quasi-quantized plateaus only form if an effective electron reservoir (such as localized states or other bands) manages to change the electronic density in the conduction as a function of magnetic field. This is of course easiest when the material's Fermi pocket is small, which again brings semimetals and semiconductors with small Fermi pockets to the center stage.

Small Fermi surfaces also present a key advantage for strongly correlated and fractionalized states in 3D because low electron densities generally favor interaction physics.[67] This can famously be illustrated by comparing the average kinetic energy $\bar{E}_{kin}$ and Coulomb energy $\bar{E}_{Coulomb}$ per electron.[68] The former is of the order of the Fermi energy, and for a parabolic band in 3D scales as $\bar{E}_{kin} \sim n^{2/3}$. A Hartree-Fock analysis furthermore shows the average Coulomb energy in 3D to scale as $\bar{E}_{Coulomb} \sim n^{1/3}$ since the density determines the average inter-electron distance.[69] Overall, the ratio of interaction energy to kinetic energy in a 3D parabolic band thus scales as $\frac{\bar{E}_{Coulomb}}{\bar{E}_{kin}} \sim n^{-1/3}$, and is larger the lower the electronic density $n$.

The importance of interactions is even larger at strong magnetic fields. Given that the motion perpendicular to the field is quenched into Landau levels, 3D materials in strong fields effectively become one-dimensional.[17,70] But one-dimensional electron systems are known to evade the Fermi-liquid paradigm, and to form strongly interacting Luttinger liquids instead.[69] This implies that interaction physics can be boosted by the application of magnetic fields of the order of the material's quantum limit – which as noted above can be reached only in materials low electronic densities.

Whenever the interaction energy dominates over both the kinetic energy and the thermal energy $k_B T$, 3D correlated ground states can be expected. This includes the possible realization of 3D analogues of the FQHE, but of course holds much more generally. Materials with small Fermi surfaces, such as topological semimetals,[71] are therefore particularly promising candidates in the quest for novel correlated phases of three-dimensional matter in magnetic fields.

***Protocol for in-situ determination of Fermi surfaces, and fingerprints of density waves.***

As explained in the above, the Hall response $\sigma_{xy} = \frac{e^2}{h} \frac{K_z}{2\pi}$ of both fully quantized and quasi-quantized quantum Hall states is characterized by a wave number $K_z$. In fully quantized states, $K_z$ is the z-component of a constant reciprocal lattice vector, while $K_z$ can vary as a function of magnetic field in quasi-quantized states. In either case, a thorough understanding of $\sigma_{xy}$ requires the identification of $K_z$ in terms of the electronic band structure.

Both the charge density wave scenario discussed in Ref. [16] as well as the quasi-quantized Hall states[17,19,54] that we focus on here relate $K_z$ to the Fermi wave number $k_{F,z}$ of the band structure at zero magnetic field. The basis for testing both scenarios is thus the *in-situ* determination of the materials' Fermi surface. Many past experiments fall short on this point.[21,23–26,29] In the following, we provide a detailed protocol to probe the morphology of the Fermi surface using SdH oscillations in $\rho_{xx}$ alongside with spectroscopic and thermodynamic experiments.

As a first crosscheck, the Hall conductivity can be used to determine if a material is actually 3D, or just layered 2D. In very anisotropic materials with small inter-layer tunneling, the Fermi surface at zero field is open. At large fields, such systems can typically be described as decoupled stacked quantum Hall layers, and $K_z = \frac{1}{a_z}$ equals the inverse lattice constant. The value of $K_z$ obtained from transport should thus in a first step be compared with the value of the inverse lattice constant determined by complementary X-ray diffraction experiments.[72]

Truly 3D materials with closed Fermi surfaces, on the contrary, exhibit quantum oscillations for magnetic fields applied along all crystal directions. Such systems can serve as parent-states for quasi-quantized states, as well as fully-quantized states resulting from either a 3D Hofstadter butterfly, or from a density wave.[16,17,19,54] For both quasi-quantized states and density wave states, the wave number $K_z = 2\, k_{F,z}$ is given by twice the zero-field Fermi-wave vector along direction at which $\vec{B}$ is applied. This Fermi wave vector can independently be determined from the SdH oscillation frequency by means of standard Landau fan diagram analysis or Fast Fourier Transforms (FFT).

In actual materials, two experimental challenges for the above protocol directly come to mind, and should be accounted for. On the one hand, the cyclotron orbit can become larger than the disorder length scale in very anisotropic crystal planes. On the other hand, if the densities are not sufficiently low and the magnetic fields not strong enough, the distance between LB edges might remain smaller than $k_B T$, which can also prevent the observation of SdH oscillations. In the past, the lack of SdH oscillations in anisotropic crystal planes has often been suggested as evidence that the Fermi surfaces of a layered materials might be quasi-2D, but in many cases a positive identification of an open Fermi surface was not provided. To avoid erroneous conclusions based solely on quantum oscillations, we propose to always corroborate the fermiology of the electronic band structures using complementary approaches. For example, the 3D carrier density relates to the Fermi wave vectors of an ellipsoidal Fermi surface *via* $n = k_{F,x} \cdot k_{F,y} \cdot k_{F,z}/(3 \cdot \pi^2)$, but also follows from the low-field slope of $\rho_{xy}$ *via* the Drude formula $n = (e \cdot d\rho_{xy}/dB)^{-1}$. Once the band structure of the material has consistently been characterized by complementary approaches, the dimensionality can be identified by comparing $k_{F,z}$ (for example obtained as $k_{F,z} = 3 \cdot \pi^2 \cdot n \cdot \hbar/(2 \cdot e \cdot f_{F,z})$ from the single SdH oscillation period of the plane perpendicular to the strong anisotropy axis) with the size of the Brillouin zone. If $k_{F,z}$ is smaller than the Brillouin zone boundary in $k_z$, the Fermi surface is closed and the system is truly 3D (instead of layered/quasi-2D). Another independent method to determine the fermiology of the Fermi surface is angle-resolved photoemission spectroscopy (ARPES).[16]

We stress that the above-described methods determine the zero-field value of $k_{F,z}$. As explained in the theory section, the Fermi wave vector will in general depend on $\vec{B}$, albeit not necessarily such that the particle number is perfectly conserved[54]. Especially Dirac and Weyl systems with low electron densities seem to have more weakly field-dependent $k_{F,z}$.[17] Still, the Hall conductivities of both quasi-quantized Hall states and field-induced density waves have been related to $k_{F,z}$ at zero field. It is thus crucial to unambiguously identify this quantity with the above recipe when analyzing $\sigma_{xy}$ in 3D systems.

The next step after the determination of a material's fermiology is to discriminate between the scenario of a field-induced charge or spin density wave with exactly quantized Hall response, and a quasi-quantized Hall effect of a 3D metal in the quantum limit. While being fundamentally very different states, both exhibit similar Hall conductivities. More clear-cut differences arise in the longitudinal resistivities. Namely, $\rho_{xx}$ is expected to vanish in the fully gapped,[5,7] truly quantized CDW/SDW scenario. In an idealized picture, this can be understood in terms of non-dissipative edge channels associated with each density wave layer. In contrast, $\rho_{xx}$ is expected to remain finite the gapless quasi-quantized state.[17,19]

Because transport in realistic samples may not follow the above idealized behavior entirely, additional measurements should also be used to corroborate the presence or absence of a CDW/SDW. In transport, the transition into a density wave state typically manifests as an abrupt increase of electrical resistance due to the gapping of the Fermi surface, and a density wave is associated with characteristic non-ohmic behavior.[73] Because non-Ohmic transport may, however, also arise in materials without a density wave,[46,74–76] complementary evidence should be collected from thermodynamic, structural, and spectral measurements. Electronic gap openings in non-magnetic materials can for example be probed by magnetization measurements,[77] thermoelectric coefficients,[78] and scanning tunneling spectroscopy.[73] The corresponding periodic charge modulation typically also leads to a change in the phonon spectrum, and reflects in Raman scattering,[73] X-ray diffraction,[79,80] and ultrasound propagation measurements.[81,82] In addition, Raman spectroscopy can directly probe Raman active CDW amplitude modes.[73] Combining such complementary evidence for or against a density wave with transport data can be very helpful to more clearly distinguish between fully and quasi-quantized states.

*Existing material candidates and ubiquity of plateau-like Hall responses in 3D.*

Several low-density materials have already been reported to exhibit plateau-like structures in their Hall response. Most prominently, this includes $ZrTe_5$[16,17], $HfTe_5$,[19,20] $SrSi_2$,[83] and InAs[54]. Some of these materials differ dramatically in their band structures, mobilities, and Fermi levels. While $ZrTe_5$ and $HfTe_5$ for example are (in parts weakly gapped) Dirac materials, InAs and $SrSi_2$ have a parabolic dispersion. As shown in Fig. 4, however, all of these materials exhibit a Hall response $\sigma_{xy} = e^2/h \cdot k_{F,z}/\pi$ once the quantum limit is reached. At the same time, $\rho_{xx}$ remains finite in all cases. We therefore propose that quasi-quantized Hall responses should be considered a natural feature of 3D metals with low electronic densities subject to strong magnetic fields. At the very least, the precise origin of 3D Hall responses in these systems, their surfaces states, the scaling relations of $\rho_{xx}$ and $\rho_{xy}$ as a function of magnetic field, and the role of correlation in these materials remains to be explored. This provides an exciting playground for future work both in theory and experiment.

# Outlook

Throughout the last decades, generalizations of the quantum Hall effect to 3D have been one of the most fruitful frontiers in the study of topology in higher dimensions. Fully gapped 3D quantum Hall states have already been proposed shortly after the discovery of the original 2D QHE, but their key signature has not been observed to date (quantized Hall plateaus combined with a vanishing longitudinal resistivity). In this perspective, we highlight gapless quasi-quantized states as an alternative. These states are consistent with quasi-quantized Hall responses alongside finite longitudinal resistances, with the precise shape of the quasi-quantized features in $\sigma_{xy}$ depending on nun-universal system specifics. We propose that quasi-quantized states may already have been realized in a multitude of distinct materials. By extension, we expect that quasi-quantized features in the Hall conductivity might generically be observed in the quantum limit of generic 3D metals and semimetals with small Fermi pockets, regardless of the precise band structure or purity. This new perspective onto Hall physics in 3D promises to explain the often puzzling plateaus appearing in Hall measurements in many other low charge carrier semimetals such as NbP[35] or HgSe[37] and in slightly doped semiconductors such as InAs[85,86] and InSb[32,87], to name a few examples.

Adding quasi-quantized states to the kaleidoscope of 3D Hall responses will contribute to future developments and lead us toward new horizons. The physics of their surface states, and in particular the interplay of these surface states with the gapless bulk states, provides an experimentally easily accessible playground for novel surface phenomena. Because the existence of a quasi-quantized plateau hinges on the presence of an effective electronic reservoir, future work should further clarify the properties of these reservoirs. Since they are likely connected to defect physics, we for example expect a correlation between sample-specific disorder properties and the visibility of quasi-quantized plateaus. The voluntary increase of defect densities during sample production would for example allow to stabilize quasi-quantized plateaus. Next, the impact of interactions in quasi-quantized and fully quantized 3D Hall states remains largely unexplored, and poses an exquisite challenge to theory. The concepts addressed in this perspective can directly be applied to also generalize the 2D quantum anomalous Hall effect[88,89] to generic 3D magnetic metals. Last but not least, quasi-quantized states also represent an important opportunity for material science: whereas conventional material science concentrated on a limited number of prominent topological materials, the stage is now open for material systems that do not belong to one of the traditional topological class, but still exhibit properties remnant of topological physics.

## Competing interests

The authors declare no competing financial interests.

## Author contributions

The authors contributed equally to all aspects of the article.

## Acknowledgments

T.M. acknowledges funding by the Deutsche Forschungsgemeinschaft via the Emmy Noether Programme ME4844/1- 1, the Collaborative Research Center SFB 1143, project A04, and the Cluster of Excellence on Complexity and Topology in Quantum Matter ct.qmat (EXC 2147). J.G. acknowledges support from the European Union's Horizon 2020 research and innovation program under Grant Agreement ID 829044 "SCHINES".

**Display items**

*Boxes*

> When a 3D material is subjected to a magnetic field $\vec{B}$, its electronic states reorganize into Landau levels. The electronic motion perpendicular to $\vec{B}$ is quenched (semiclassically, the electrons are confined to cyclotron orbits), but the Landau levels still disperse with the momentum $k_\parallel$ parallel to $\vec{B}$. For $\vec{B} = B\,\vec{e}_z$, the material can be thought of as Landau levels in the $(x,y)$-plane stacked along $k_z$. The 3D band structure is thus reorganized into a set effectively one-dimensional Landau bands $j$ with energies $E_j(k_z)$. As usual for Landau levels, the Landau bands are macroscopically degenerate. The total Hall response $\sigma_{xy}$ is given by the integral over $k_z$ of the Hall responses of all effective 2D subsystems defined by a given $k_z$,
>
> $$\sigma_{xy} = \sum_j \frac{1}{2\pi} \int \sigma_{xy,j}(k_z)\, dk_z.$$
>
> If for a given momentum $k_z$, the energy $E_j(k_z)$ of the j-th Landau band is below the Fermi level, then $\sigma_{xy,j}(k_z) = \frac{e^2}{h}$. If on the contrary $E_j(k_z)$ is above the Fermi level, then $\sigma_{xy,j}(k_z) = 0$.
>
> When the only the zeroth Landau level is occupied, the total Hall conductivity is thus proportional to the occupied momentum range in the zeroth Landau band,
>
> $$\sigma_{xy} = \frac{1}{2\pi}\int \sigma_{xy,0}(k_z)\, dk_z = \frac{1}{2\pi}\frac{e^2}{h}\int_{occ} dk_z = \frac{1}{2\pi}\frac{e^2}{h} 2\, k_{Fz,0},$$
>
> where $k_{Fz,0}$ is the Fermi momentum in the zeroth Landau band.

Box 1 | **Theory: Scaling relation of the Hall effect in three dimensions.**

*Figures*

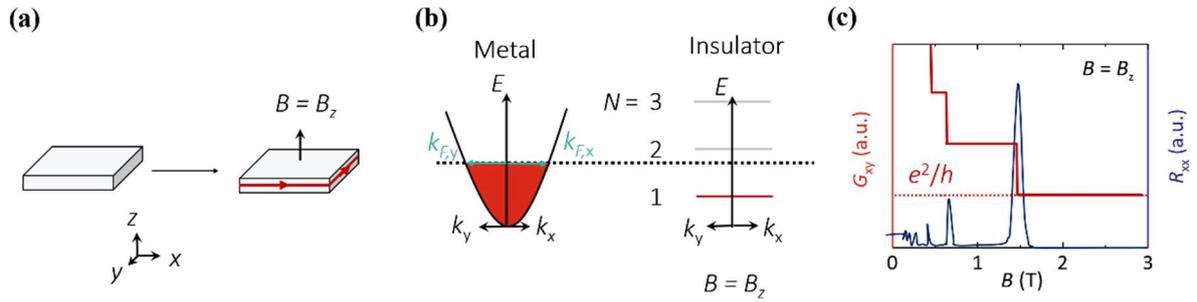

Figure 1 | **The quantum Hall effect in a 2D electron system. a** | Sketch of a 2D metal without (left) and with a magnetic field (right) applied in the z-direction ($B = B_z$). **b** | Sketch of a 2D band structure without (left) and with a strong $B = B_z$ applied (right). The $k_i$ anf $k_{F,i}$ refer to the momentum space components and the Fermi wave vectors in $i = x, y$-direction, respectively. $N$ denotes the Landau level index and $E$ the energy. **c** | Sketch of the Hall conductance $G_{xy}$ (left axis, red) and longitudinal resistance $R_{xx}$ (right axis, blue) as a function of $B$.

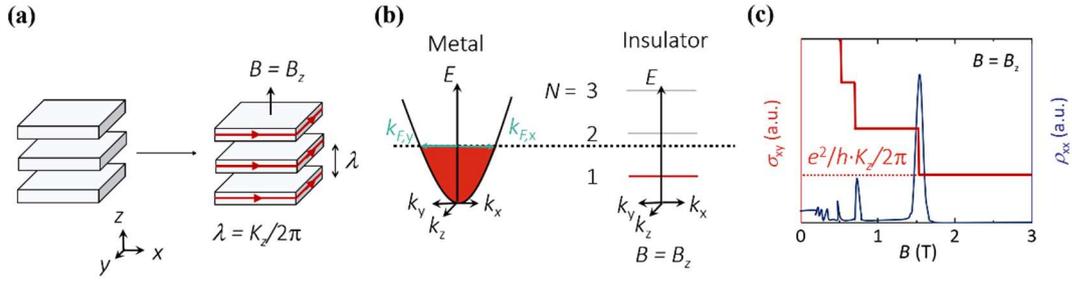

Figure 2 | **The quantum Hall effect in a 2D multilayer system. a** | Sketch of a 2D multilayer metal without (left) and with a magnetic field (right) applied in the z-direction ($B = B_z$). $K_z$ is the reciprocal lattice vector component along $B$ that corresponds to the periodicity $\lambda$ of the multilayer system. **b** | Sketch of the band structure without (left) and with a strong $B = B_z$ applied (right) in a single 2D layer of a multilayer system. The $k_i$ and $k_{F,i}$ refer to the momentum space components and the Fermi wave vectors in $i = x, y, z$-direction, respectively. N denotes the Landau level index and E the energy. **c** | Sketch of the total Hall conductivity $\sigma_{xy}$ (left axis, red) and longitudinal resistivity $\rho_{xx}$ (right axis, blue) as a function of $B$.

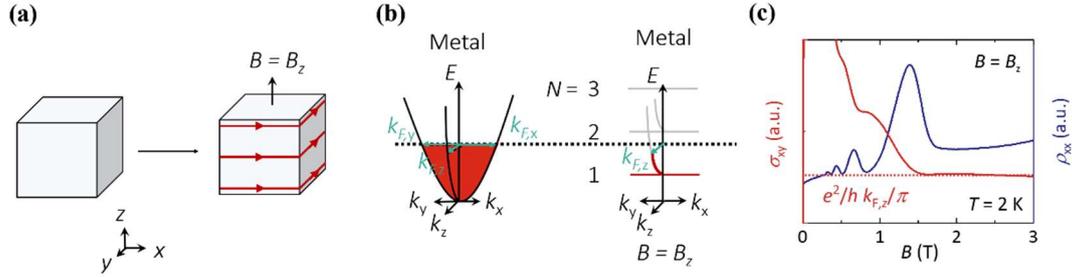

Figure 3 | **Quantum Hall-like physics in three dimensions. a** | Sketch of a 3D system without (left) and with a magnetic field applied in the z-direction ($B = B_z$). b) Sketch of a 3D band structure without (left panel) and with a strong $B = B_z$ applied. The $k_i$ and $k_{F,i}$ refer to the momentum space components and the Fermi wave vectors in $i = x, y, z$-direction, respectively. $N$ denotes the Landau band index and E the energy. c) Hall conductivity $\sigma_{xy}$ (left axis, red) and longitudinal resistivity $\rho_{xx}$ (right axis, blue) of a ZrTe$_5$ sample as a function of $B$ applied the z-direction at 2 K. The data is taken from Ref [17].

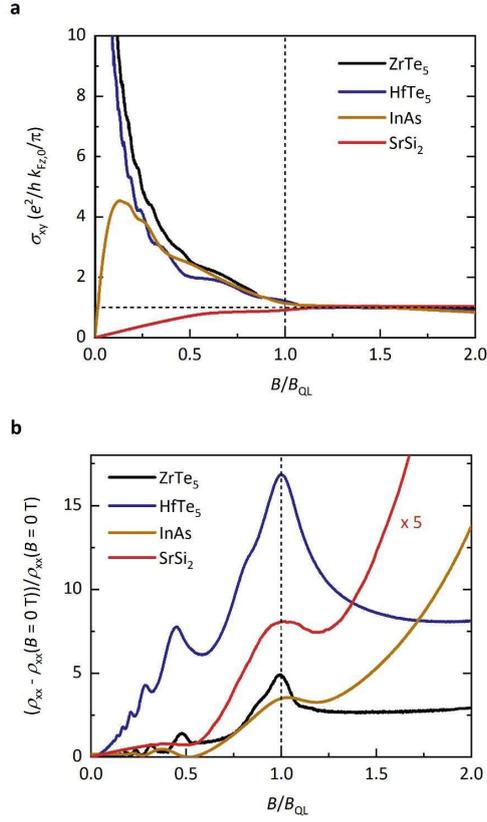

Figure 4 | **Similarities of the Hall effect in the quantum limit of 3D metals. a** | In spite of large variations in their Fermi wavevectors in the lowest Landau band ($k_{F,z,0}$) along the applied magnetic field $B$, a wide range of metals exhibit a $e^2/h \cdot k_{F,z,0}/\pi$-scaling of the Hall conductivity $\sigma_{xy}$ in the quantum limit $B > B_{QL}$ (ZrTe$_5$:[17] $B_{QL}$ = 1.2 T, $k_{F,z,0}$ = (79 × 10$^{-3}$) Å$^{-1}$; HfTe$_5$:[36] $B_{QL}$ = 1.8 T, $k_{F,z,0}$ = (58 × 10$^{-3}$) Å$^{-1}$; InAs:[54] $B_{QL}$ = 1.9 T, $k_{F,z,0}$ = ((8 × 10$^{-3}$) Å$^{-1}$–$e \cdot B/\hbar$)$^{1/2}$ Å$^{-1}$; SrSi$_2$:[83] $B_{QL}$ = 11.4 T, $k_{F,z,0}$ = ((20 × 10$^{-3}$) Å$^{-1}$–$e \cdot B/\hbar$)$^{1/2}$ Å$^{-1}$). **a** | The longitudinal resistivity $\rho_{xx}$ normalized by its zero-field value $\rho_{xx}$ ($B$ = 0 T). $\rho_{xx}$ remains always finite for all materials investigated. All data is taken at a temperature of around 2 K. The universality of the $k_{F,z}$-scaling is observed in spite of the fact that the band structures vary across the range of materials. ZrTe$_5$, and HfTe$_5$ exhibit a Dirac-type lowest Landau band, while InAs and SrSi$_2$ exhibit lowest Landau bands with a parabolic dispersion along $B$.